\documentclass[preprint]{revtex4}

\usepackage{amsfonts,amsmath,amsxtra,graphicx} 

\begin{document}

\title{Temporal contribution to gravitational WKB-like calculations}

\author{Valeria Akhmedova}
\email{lera@itep.ru}
\affiliation{Moscow, B.Cheremushkinskaya, 25, ITEP, Russia 117218} 

\author{Terry Pilling}
\email{Terry.Pilling@ndsu.edu}
\affiliation{Department of Physics,
North Dakota State University, Fargo, ND 58105} 

\author{Andrea de Gill}
\email{aadegill@csufresno.edu}
\author{Douglas Singleton}
\email{dougs@csufresno.edu}
\affiliation{Physics Department, CSU Fresno, Fresno, CA 93740-8031} 

\date{\today} 

\begin{abstract}
Recently, it has been shown that the radiation arising from
quantum fields placed in a gravitational background (e.g. Hawking
radiation) can be derived using a quasi-classical calculation.
Here we show that this method has a previously overlooked temporal
contribution to the quasi--classical amplitude. The source of this
temporal contribution lies in different character of time in
general relativity versus quantum mechanics. Only when one takes
into account this temporal contribution does one obtain the
canonical temperature for the radiation. Although in this letter the
specific example of radiation in de Sitter space--time is used, the
temporal contribution is a general contribution to the
radiation given off by any gravitational background where the time
coordinate changes its signature upon crossing a horizon. Thus,
the quasi--classical method for gravitational backgrounds contains
subtleties not found in the usual quantum mechanical tunneling
problem.
\end{abstract} 

\maketitle

\section{Introduction} 

One of the consequences of placing quantum fields in a gravitational
background is that these backgrounds can emit radiation. Examples
include Hawking radiation \cite{hawking} for black holes, Unruh
radiation \cite{unruh} for an accelerated observer, and Gibbons-Hawking
radiation \cite{gibbons}
for an observer in de Sitter space. In each of these cases the
spectrum of the radiation is thermal or Planckian. Also each of these 
space--times
has at least one horizon which separates the space--time into different
sections. 

There are different methods of calculating the radiation from a
gravitational background. One method uses the fact that the
annihilation/creation operators for some quantum field are in
general different in the different space--time sections. By
finding the Bogolubov transformation between the different sets of
annihilation/creation operators, the spectrum of the
radiation emitted can be obtained. Another method uses Wightman functions to 
study
the response of a detector to a given gravitational background
\cite{birrel} (see as well \cite{cir-unruh}). 

In \cite{kraus} \cite{parikh} a method was developed where
the Hawking radiation from a Schwarzschild
black hole was obtained using a quasi--classical method. (However
see \cite{brout} for an earlier quasi--classical calculation of
Gibbons-Hawking radiation for de Sitter space--time). The
quasi--classical amplitude is given by the exponent of the imaginary
part of the classical action for the particles coming from the near
horizon region. There are many advantages to this quasi-classical
calculation: (i) The calculations are simple; (ii) One can apply
this method to a host of gravitational backgrounds and different
spin particles; (iii) The calculation gives a microscopic picture of
the radiation; (iv) The back--reaction of the radiation on the black
hole metric may be taken into account \cite{parikh};  (v) This
method may also be used to obtain the standard relationship between
the black hole temperature and entropy \cite{pilling}. 

However, the interpretation of the imaginary
contribution to the particle's action as an indication of
tunneling has some subtleties. First, if the pair is
created behind the horizon neither of the particles can tunnel
through the horizon, because the tunneling process in quantum
mechanics is described via the solution of a Cauchy problem and has
to be causal, while passing through the horizon is acausal. In
quantum mechanics the vacuum remains unchanged
which is the reason why we can safely convert a time evolution problem into
an eigen--value problem.
Second, if the pair is created outside a horizon
the time for one of the particles to cross the horizon is
infinite for the stationary distant observer.
However, this same observer should see the radiation
from the black hole in finite time after the collapse.
In \cite{parikh} the above subtleties were addressed by taking the
horizon to shrink during the pair creation process so that the
radiated particles appeared already outside the horizon. 

In this letter the quasi--classical picture is applied to de Sitter
space--time using the Hamilton--Jacobi equations. Using the
Hamilton--Jacobi equations to find the imaginary contribution to
the classical action in a gravitational background is analogous to
the Trace-Log calculations in quantum field theory, e.g. finding
the probability of vacuum decay in an external electro--magnetic
field \cite{schwinger}. In a Trace-Log calculation one would look
for the imaginary contribution to the vacuum decay amplitude, i.e.
Tr~$\log{[\Box(g) + m^2]} = \int Dx(t) \, e^{-\frac{i}{\hbar}\,
S(g,x)}$. $\Box(g)$ is the d'Alembertian operator in the
background metric $g_{\mu \nu}$ for a particle of mass $m$. On the
right hand side of the equation is a path integral over closed paths
and $S$ is the action for particles in the gravitational field. In
the quasi--classical approximation ($\hbar \to 0$) the
saddle--point approximation for the path integral is used. The imaginary
contribution comes from the closed paths which cross the horizon
going out and back. Similarly in quasi--classics with the
Hamilton--Jacobi equations, the imaginary contribution is found from
an integral which spans the horizon. 

\section{Hamilton-Jacobi Equations} 

For a scalar field, $\phi (x) \propto \exp\{ -\frac{i}{\hbar} S(x) + ...
\}$, of mass $m$, the Hamilton-Jacobi equations (to $0^{th}$ order in 
$\hbar$) are
\begin{equation}
\label{hj}
g^{\mu \nu} (\partial _\mu S ) (\partial _\nu S) + m ^2 = 0 ~.
\end{equation}
$S(x)$ is the action of the scalar field and $g^{\mu \nu}$ is
the metric of the background space--time. 

For stationary space--times with a time--like Killing vector the
action can be expressed as the sum of the time and spatial parts
\begin{equation}
\label{phase}
S (x ^\mu ) = Et + S_0 (\vec{x}),
\end{equation}
where $E$ is the particle energy and $x^\mu = (t, \vec{x})$.
Inserting some particular metric into (\ref{hj}) gives an equation
for $S_0 ( \vec{x})$ which has the solution $S_0 = - \int p_r dr$
where $p_r$ is the canonical momentum for the metric. If $S_0$ has
an imaginary part the temperature of the
radiation can be obtained as follows: the decay rate due to the quasi--
classical
\begin{equation}
\label{gamma}
\Gamma \propto  \exp \left[ - {\rm Im} \oint p_r dr / \hbar \right]
= \exp \left[  - {\rm Im} \left( \int p_r ^{Out} dr - \int p_r ^{In} dr
\right) /
\hbar \right] .
\end{equation}
The closed path of integration in the first expression goes across
the barrier (i.e. the horizon) {\it and} back. The temperature of the 
radiation is
obtained by associating the expression in \eqref{gamma} with a
Boltzmann factor $\Gamma \propto \exp [ -E /T ]$ which gives $T
= \frac{\hbar E}{{\rm Im} \oint p_r dr}$ (note that $T$ is independent
of $E$ and we have set $k_B =1$). 

Sometimes the decay rate in \eqref{gamma} is written as $\Gamma
\propto \exp [ \pm 2 {\rm Im} \int p_r ^{Out, In} dr / \hbar ]$
since in many cases $p_r ^{Out} = - p_r ^{In}$ i.e. crossing the
horizon left to right versus right to left only differs by a sign.
However, it has been shown \cite{chowdhury} that $\exp [ \pm 2 {\rm
Im} \int p_r ^{Out, In} dr / \hbar ]$ is not invariant under
canonical transformation but that $\oint p_r dr$ is invariant.
Thus, only $\oint p_r dr$ is a proper observable and only $\oint p_r dr$
should be used in \eqref{gamma}. In the standard
quantum mechanical tunneling problem it does not matter whether
one uses $\oint p_r dr$ or $\exp [ \pm 2 {\rm Im} \int p_r ^{Out,
In} dr / \hbar ]$ in the exponent of \eqref{gamma} since the
tunneling is the same independent of the direction in which one
crosses the barrier i.e. $p_r ^{Out} = - p_r ^{In}$. In the
gravitational quasi--classical problem there are cases in which
the passing across the horizon {\it does} depend on the direction
of traversal i.e. $p_r ^{Out} \ne - p_r ^{In}$. An example of this
occurs for the black hole solution in Painlev{\'e} coordinates
\cite{akhmedov}. 

\section{Spatial Contribution}
Using the de Sitter space--time the
``spatial" contribution to $\Gamma$ is calculated, i.e. the
contribution coming from the imaginary part of $S_0 (\vec{x})$.
The full 4D de Sitter
space--time can be seen as a hyperboloid $- z_0 ^2 + z_1 ^2 + z_2
^2 + z_3 ^2 + z_4 ^2 =  \alpha ^2$ embedded in 5D Minkowski
space--time $ds^2 = - dz_0 ^2 + dz_1 ^2 + dz_2 ^2 + dz_3 ^2 + dz_4
^2$. Transforming to static coordinates
\begin{eqnarray}
\label{static}
z_0 &=& (\alpha ^2 -r^2)^{1/2} \sinh (t/\alpha) ~~,~~~~~~  z_1 = (\alpha ^2
 -r^2)^{1/2} \cosh (t/\alpha) \nonumber \\
z_2 &=& r \sin \theta  \cos \phi  ~~, ~~~~~~ z_3 = r \sin \theta  \sin \phi
~~,~~~~~~ z_4 = r \cos \theta  ~~,
\end{eqnarray}
the de Sitter space--time takes the Schwarzschild--like form
\begin{equation}
\label{desitter}
ds^2 = - \left( 1-\frac{r^2}{\alpha^2} \right) dt^2 + \frac{dr^2}{\left(
1-\frac{r^2}{\alpha^2} \right)}
 + r^2 ( d \theta^2 + \sin ^2 \theta  d \phi ^2 )
\end{equation}
This form of the de Sitter metric has a coordinate singularity at
$r = \alpha$ which is the event horizon.

Using metric \eqref{desitter} in \eqref{hj} the following
solution for $S_0$ is found
\begin{equation}
\label{sol1}
S_0 ^{In, Out} = \pm \int_0^{+\infty} \frac{\sqrt{E^2 - m^2 \left(1 -
\frac{r^2}{\alpha ^2} \right)}}{ 1- \frac{r^2}{\alpha ^2}} dr =
\pm \int_0^{+\infty} \frac{\alpha ^2\sqrt{E^2 - m^2 \left(1 -
\frac{r^2}{\alpha ^2} \right)}}{ (\alpha - r)(\alpha + r)} dr .
\end{equation}
Because of spherical symmetry the angular part can be neglected.
The $+(-)$ sign correspond to different traversal directions
across the horizon at $r=\alpha$ i.e. $+=In$ and $-=Out$. There is a pole
on the $r$-axis at $r=\alpha$. One can evaluate the imaginary part of
\eqref{sol1} using a semi-circular contour. The result is $S_0 ^{In, Out}
= \pm \frac{i \pi E \alpha}{2}$.
The total contribution coming from the spatial part of the action (i.e. from
$S_0$) is obtained by using the closed path which goes
out and back across the horizon. The difference in the sign
of $S_0 ^{In, Out}$ is then compensated by a reversal of the integration
path with the result that the exponent in \eqref{gamma} is
$\oint p_r dr = + i \pi E \alpha$. This yields a temperature of $T =
\frac{\hbar}{\pi \alpha}$ which is
twice the Gibbons-Hawking temperature. This factor of two greater
temperature was already noticed in \cite{brout}. One might suspect that the
static coordinates in \eqref{static} are ``bad" since they do
not cover the entire de Sitter space--time, and that one should use another
coordinate system which does cover the entire space--time.
However since $\oint p_r dr$ is invariant under canonical transformations
this result will remain the same in any other coordinate system related
to \eqref{desitter} by a canonical transformation. 

Below it is shown that this disagreement comes from a
missed contribution from the temporal part of the total action, $S (x 
^\mu)$. 

\section{Temporal Contribution} 

The static coordinates \eqref{static} cover only the right quadrant
of the Penrose diagram of the de Sitter space--time. On crossing the
horizon at $r=\alpha$ the coordinates $z_0$ and $z_1$ reverse their
time-like/space-like character. Thus, $z_0 \propto \cosh(t/\alpha) $
and $z_1 \propto \sinh (t/\alpha)$. Also now $r
> \alpha$ so the order under the square root should be reversed.
Therefore the left quadrant of the Penrose diagram
of the de Sitter space--time is covered by the
following coordinates
\begin{eqnarray}
\label{static1}
z_0 &=& (r^2 - \alpha ^2)^{1/2} \cosh (t/\alpha) ~~,~~~~~~  z_1 = (r^2 -
\alpha ^2)^{1/2} \sinh (t/\alpha) \nonumber \\
z_2 &=& r \sin \theta  \cos \phi  ~~, ~~~~~~ z_3 = r \sin \theta  \sin \phi
~~,~~~~~~ z_4 = r \cos \theta  ~~,
\end{eqnarray}
The coordinates in \eqref{static} are related to those in
\eqref{static1} simply by letting $t \rightarrow t - i\frac{\pi
\alpha}{2}$ in \eqref{static}. Under this transformation
$\sinh(t/\alpha) \rightarrow -i \cosh (t /\alpha)$ and
$\cosh(t/\alpha) \rightarrow -i \sinh (t /\alpha)$. The $-i$ is
taken care of by an $i$ coming from the square root part. Thus, on
crossing the horizon there is a temporal contribution to the
imaginary part of the total action, $S(x_\mu)$, given by ${\rm
Im}(E \Delta t ^{In, Out}) = - \frac{\pi \alpha E}{2}$. 

Thus, the total imaginary part of the full action (i.e. $S (x ^\mu
) = Et + S_0 (\vec{x})$) has equal magnitude contributions from
the time and spatial parts. Explicitly (remebering that
$S_0 = - \int p_r dr$) we have
${\rm Im} (S (x ^\mu )) = {\rm Im}(E
\Delta t^{Out} + E\Delta t ^{In} - \oint p_r dr ) = - \frac{\pi \alpha
E}{2} - \frac{\pi \alpha E}{2} - \pi \alpha E = - 2 \pi \alpha E$. 
This yields the canonical Gibbons-Hawking
temperature of $T_{GH} =  \frac{\hbar}{2 \pi \alpha}$. 

\section{Conclusion} 

Using de Sitter space--time it has been shown that the quasi-classical
calculations have an additional subtle feature relative to the
standard quantum mechanical tunneling problem.  In addition to the
spatial contribution to the quasi--classical amplitude (i.e. ${\rm Im}
(S_0)$) we have shown that there is an equal contribution coming
from the temporal part (i.e. ${\rm Im} (E \Delta t)$. This is not
unexpected since time is treated differently in
general relativity and quantum mechanics.
In the former it is a dynamical coordinate
like the spatial coordinates; in the latter it is a parameter
which is distinct from the spatial coordinates.
Although the specific case of de Sitter space--time was studied, the
same temporal contribution will occur when one crosses a horizon where
the time and spatial coordinates switch their time-like/space-like
character upon crossing the horizon. E.g. similar temporal contribution 
appears in the
Schwarzschild space--time which exactly fixes the factor of two
problem in the Hawking temperature remarked on in \cite{akhmedov}
(see \cite{AkPiSi}). Other recent work dealing with the Hawking-Gibbons 
radiation of de Sitter spacetime via tunneling can be found in \cite{volovik}.

Rindler space--time and the associated Unruh radiation provides an
example where one has a horizon but does not have a temporal
contribution to the quasi--classical amplitude. Rindler
space--time for an acceleration $a$ is given by
\begin{equation}
\label{Rind}
ds^2 = - (1+a\,x)^2\, dt^2 + dx^2 + dy^2 + dz^2.
\end{equation}
One can see that the $t$ coordinate remains time-like on crossing
the horizon at $x=-1/a$. Thus Unruh radiation does not receive a
temporal contribution and one obtains the canonical Unruh
temperature from only the spatial contribution to the
quasi--classical amplitude (see the second reference in
\cite{akhmedov}). 

\begin{center}
{\bf Acknowledgments}
\end{center}
Authors would like to thank E.T.Akhmedov for valuable discussions.
VA would like acknowledges support from the Agency of Atomic
Energy of Russian Federation.


\end{document}